\documentclass[aip,jcp,reprint,amsmath,amssymb,superscriptaddress]{revtex4-2}

\usepackage{graphicx}
\usepackage{booktabs}
\usepackage{bm}
\newcommand{\tcell}[2]{\parbox[t]{#1}{\raggedright\strut #2\strut}}

\newcommand{\Smite}{\textsc{Smite}}
\newcommand{\kb}{k_{\mathrm B}}
\newcommand{\Ecoll}{E_{\mathrm{coll}}}
\newcommand{\bmax}{b_{\max}}
\newcommand{\code}[1]{\texttt{#1}}
\usepackage{xcolor}
\usepackage{hyperref}
\hypersetup{colorlinks=true,linkcolor=blue!50!black,urlcolor=blue!50!black,citecolor=blue!50!black}

\begin{document}

\title{Smite: A quasiclassical trajectory (QCT) program for bimolecular collisions and unimolecular dynamics on ab-initio, and machine-learned potential energy surfaces}

\author{P\'eter Szab\'o}
\email{peter88szabo@gmail.com}
\email{peter.szabo@kuleuven.be}
\affiliation{Belgian Institute for Space Aeronomy (BIRA-IASB), Brussels, Belgium}
\affiliation{Department of Chemistry, KU Leuven, Leuven, Belgium}
\author{Jenne van Veerdeghem}
\affiliation{Department of Chemistry, KU Leuven, Leuven, Belgium}
\author{J\'er\^ome Loreau}
\affiliation{Department of Chemistry, KU Leuven, Leuven, Belgium}

\author{Jean-Fran\c{c}ois M\"uller}
\affiliation{Belgian Institute for Space Aeronomy (BIRA-IASB), Brussels, Belgium}
\author{Jeremy N. Harvey}
\affiliation{Department of Chemistry, KU Leuven, Leuven, Belgium}

\date{\today}

\begin{abstract}
We present \Smite, a Python toolkit for quasiclassical trajectory (QCT) simulations of reactive and inelastic bimolecular collisions, unimolecular dynamics, and molecular collisions with finite surface models. A central feature is the explicit control of reactant preparation: harmonic normal modes admit fixed quantum numbers, prescribed energies, thermal quantum populations, or ground-state Wigner sampling; rotation is prepared at fixed angular-momentum magnitude or from thermal distributions appropriate to diatomic and polyatomic rotors. Rotating-Morse sampling provides a coupled anharmonic treatment of diatomic vibration and rotation, while saved molecular-dynamics phase points provide an alternative source of initial conditions. The same nuclear propagators operate with on-the-fly electronic-structure calculations or user-supplied analytical and machine-learned potential energy surfaces. Additional modules provide constrained rigid-fragment dynamics, thermostat-based preparation, photoionization initial conditions with energy and recoil constraints, and fewest-switches surface hopping (FSSH) on multiple PESs with approximate curvature-derived couplings. Geometry optimization, transition-state searches, crossing-point optimization, intrinsic-reaction-coordinate following, and thermochemistry use the same energy and derivative interfaces. Analysis tools connect trajectories to product energy partitioning, stereodynamical correlations, time-resolved vibrational signatures, and independent-atom x-ray scattering. \end{abstract}

\maketitle
\raggedbottom

\section{Introduction}
\label{sec:intro}

Reaction dynamics experiments and modeling have been at the focus of chemists' attention since the 1930s, because they provide information about the course of chemical reactions at the atomic level. The discovery of the potential barrier on the H\,$+$\,H$_2$ potential energy surface by Eyring and M.~Polanyi \cite{EyringPolanyi1931} shed light on the origin of the activation energy that Arrhenius had introduced decades earlier without a convincing microscopic explanation. Polanyi and co-workers \cite{EvansPolanyi1939} also strived to understand why some reactions produce vibrationally excited products whereas others do not---questions that culminated, decades later, in the celebrated Polanyi rules connecting the location of the barrier to the efficacy of translational versus vibrational excitation \cite{Polanyi1972}. The conceptual foundation of essentially all of this work is the Born--Oppenheimer approximation: the separation of electronic and nuclear motion. The solution of the electronic Schr\"odinger equation provides a stationary, multidimensional potential energy surface (PES) that formally supplies the forces governing the motion of the nuclei. Once the PES is known, ``all'' that remains is to solve the nuclear equations of motion---preferably quantum mechanically or, relying on the relatively large mass of the nuclei, classically. The technology for the numerical simulation of molecular collisions using classical mechanics has in principle been available since Newton, Lagrange, Euler, and Hamilton; the missing ingredients were the computational power and the potential energy surface.

The first classical trajectory simulations of unimolecular \cite{Bunker1962} and bimolecular \cite{BlaisBunker1962,KarplusPorterSharma1965} reactions therefore relied on empirical potential functions with adjustable parameters, most prominently London--Eyring--Polanyi--Sato (LEPS) surfaces for atom\,$+$\,diatom systems. One of the purposes of such calculations, exemplified by the work led by J.~C.~Polanyi, was precisely to derive qualitative rules governing chemical reactions by adjusting the topography of the surface at will. With the advent of accurate ab initio electronic-structure methods, globally fitted potential energy surfaces of near-spectroscopic quality became available for small systems---the F\,$+$\,H$_2$ surface of Stark and Werner \cite{StarkWerner1996} being an early landmark---and the emphasis of the field shifted toward quantitative agreement between dynamics calculations and experiment, including full-dimensional quantum dynamics for systems as large as H\,$+$\,CH$_4$ \cite{WelschManthe2015}. High-quality fitted surfaces such as those for HO$_2$ in its ground \cite{Xu2005} and first excited \cite{Li2010} electronic states enabled detailed comparative studies of complex-forming reactions \cite{Lendvay2008,Sun2008,Guo2012,SzaboLendvay2015a,SzaboLendvay2015b}, and QCT calculations on such surfaces remain one of the most productive tools of gas-phase reaction dynamics \cite{TruhlarMuckerman1979,RaffThompson1985,AoizBanaresHerrero1998}.

Two developments over the past three decades have qualitatively changed how the PES enters a trajectory calculation. The first is the emergence of systematic, high-dimensional \emph{fitting} strategies. Modified Shepard interpolation \cite{Collins2002}, permutationally invariant polynomial (PIP) expansions \cite{BraamsBowman2009,QuYuBowman2018}, and, more recently, machine-learning (ML) representations---high-dimensional neural-network potentials \cite{BehlerParrinello2007,Behler2016}, PIP--neural-network hybrids \cite{JiangGuo2013}, Gaussian-approximation and kernel methods \cite{Bartok2010}, and message-passing architectures such as PhysNet \cite{UnkeMeuwly2019} or transferable models like ANI \cite{Smith2017}---make it possible to represent reactive PESs of systems with ten and more atoms with close to ab initio accuracy at a fraction of the cost \cite{ManzhosCarrington2021,Meuwly2021}. The second development is the maturation of \emph{direct dynamics}, in which the potential energy and its gradient are obtained on the fly from an electronic-structure calculation at every step of the trajectory, without any intermediate analytical fit \cite{WangKarplus1973,Leforestier1978,BoltonHasePeslherbe1998,SunHase2003,Paranjothy2013}. Direct dynamics is the more general approach, because constructing an analytical or ML function that accurately represents a chemical process with multiple pathways, transition states, and minima remains a formidable and lengthy task; its price is the large number of electronic-structure calculations required per trajectory. Considerable algorithmic effort has therefore been devoted to enhancing the efficiency of direct dynamics, including Hessian-based predictor--corrector integrators \cite{Millam1999,Lourderaj2007}, Hessian-updating schemes \cite{Bakken1999,Bofill1994}, and high-accuracy compact-finite-difference Hessian approximations \cite{Wu2010}. Born--Oppenheimer direct dynamics converges the electronic problem at each geometry, whereas extended-Lagrangian Car--Parrinello dynamics propagates auxiliary electronic degrees of freedom \cite{CarParrinello1985}. Their relative cost and accuracy depend on the system, electronic-structure model, and numerical settings; in either case the quality of the nuclear dynamics is limited by the underlying electronic approximation. Practical direct-dynamics packages typically couple a chemical dynamics driver with one or several quantum-chemistry engines---the archetype being VENUS \cite{Hase1996} interfaced with Gaussian, GAMESS, NWChem, or MOPAC \cite{Paranjothy2013}---and simulations of this kind have repeatedly uncovered qualitatively new mechanistic phenomena, such as the roundabout mechanism of the Cl$^-$\,$+$\,CH$_3$I S$_\mathrm{N}$2 reaction \cite{Mikosch2008}, reactions that avoid their deep potential energy minima \cite{Sun2002}, and dominant non-IRC reaction paths \cite{Lopez2007}.

Against this historical background, the following subsections discuss when explicit dynamics is needed, the collision processes it can describe, and the scope of the present work.

\subsection{Why explicit dynamics? Statistical theories and their limits}
\label{sec:whydynamics}

A large part of chemical kinetics is built on statistical rate theories, but their physical domains and assumptions should be distinguished. For reactions with a well-defined potential-energy barrier, canonical and microcanonical (variational) transition state theory (TST) provide established descriptions of the rate when the reactant population is appropriately represented and recrossing is negligible or corrected for \cite{Eyring1938,Wigner1938,BaoTruhlar2017}. Barrierless association instead involves capture governed by long-range attraction, centrifugal barriers, and the anisotropy of the interaction. Phase space theory \cite{Light1964,Nikitin1974}, orbiting transition state theory \cite{ChesnavichBowers1977}, and the statistical adiabatic channel model (SACM) \cite{QuackTroe1974,Troe1983,TroeUshakov2006} provide descriptions of capture and the reverse dissociation process with different treatments of the accessible channels. Variable-reaction-coordinate TST extends the dividing-surface approach to such loose entrance channels \cite{GeorgievskiiKlippenstein2003a,GeorgievskiiKlippenstein2003b,MillerKlippenstein2004,Zhang2020}. Predicting capture does not by itself require complete intramolecular energy redistribution; a statistical description of the subsequent decay and branching additionally requires sufficient loss of memory within the collision complex. Its applicability depends on the well depth and anisotropy, collision energy, and the competition between redistribution and escape; a deep well alone does not guarantee statistical behavior \cite{Guo2012}. Likewise, RRKM theory for unimolecular reactions assumes sufficiently rapid intramolecular energy redistribution \cite{BaerHase1996}. Explicit dynamics can test the relevant assumptions by retaining the time evolution and correlations of individual encounters.

These assumptions must therefore be assessed for the particular reaction and collision conditions. Defining and numerically locating a one-dimensional reaction coordinate is far from straightforward for loose, barrierless entrance channels, and the difficulty is amplified for flexible molecules in which multiple conformers interconvert on the timescale of the collision as the reactants approach each other on a flat PES \cite{Miller1976a,Miller1976b}. The transitional degrees of freedom can be strongly coupled, so that simple normal-mode analysis is inadequate, and flexible or variable-reaction-coordinate transition state theories can require expensive Monte Carlo sampling of the transitional phase space with on-the-fly energy evaluations \cite{RobertsonWagnerWardlaw2000,GeorgievskiiKlippenstein2003a}. More fundamentally, explicit trajectory studies have documented numerous systems in which complex decay does not follow statistical predictions. For the complex-forming H\,$+$\,O$_2$($^3\Sigma_g^-$) and H\,$+$\,O$_2$($^1\Delta_g$) reactions---prototypes of combustion chain branching---the lifetime distributions of the HO$_2$ collision complexes are strongly nonexponential, reactive and nonreactive decay channels do not obey statistical branching, product angular distributions retain the memory of the initial approach geometry, and the reactivity is governed by isomerization dynamics within the well rather than by state counting \cite{SzaboLendvay2015a,SzaboLendvay2015b,Lendvay2008,Sun2008,Lin2006,Bargueno2007}. For the reverse O\,$+$\,OH\,$\rightarrow$\,H\,$+$\,O$_2$ reaction, comparisons of QCT, quantum scattering, and statistical calculations reveal both statistical and nonstatistical aspects, illustrating that the adequacy of a statistical description can depend on the observable considered \cite{Jorfi2009}. Gas-phase S$_\mathrm{N}$2 reactions display nonstatistical central-barrier recrossing, mode-specific dynamics, and nontraditional atomistic mechanisms \cite{Cho1992,Sun2002,Mikosch2008,Manikandan2012}. Explicit dynamics also provides information beyond thermal rate coefficients for reactions with barriers: Czak\'o and Bowman showed for Cl\,$+$\,CHD$_3$ that, at low collision energies, CH-stretch excitation is no more effective than an equivalent increase in translational energy, despite the late barrier \cite{CzakoBowman2011}. Such state-specific behavior need not imply a failure of TST for the thermal rate. For barrierless radical--radical or radical--O$_2$ associations, where capture theories are the standard tool, the outcome can be controlled by orientational and conformational dynamics during the approach: in our recent on-the-fly QCT study of O$_2$ addition to the chemically activated $Z,Z'$-OH-allyl radicals central to isoprene oxidation, the site-specific ($\alpha$/$\gamma$) branching of the barrierless capture---a quantity that challenges simple statistical models because of the strongly coupled, conformationally flexible transitional modes---was determined directly from the collision dynamics \cite{Szabo2025capture}. Explicit trajectory simulation is therefore particularly useful when the timescales of intramolecular energy redistribution, conformational motion, and reaction are not cleanly separated; it requires no predefined reaction coordinate, no separability, and no ergodicity assumption, and it delivers, in addition to rate coefficients, the full mechanistic movie of the process \cite{Peslherbe1999,Paranjothy2013,Lourderaj2008}.

\subsection{Collision processes across chemical environments}
\label{sec:environments}

The processes amenable to QCT simulation span essentially every environment in which molecules collide. In \emph{combustion}, the H\,$+$\,O$_2 \rightarrow$ O\,$+$\,OH reaction and its reverse constitute the single most important chain-branching step \cite{Miller1990,Miller2005}, and both its reactive dynamics \cite{Lendvay2008,SzaboLendvay2015b} and the highly efficient nonreactive vibrational excitation of O$_2$ in ``frustrated'' complex-forming collisions \cite{SzaboLendvay2015b} have been characterized by trajectories; QCT calculations also provide high-temperature rate coefficients for reactions that are difficult to access experimentally, such as CH$_3$\,$+$\,HBr \cite{Goger2018,Gao2022,Bedjanian2023,SzaboLendvay2024}. For OH\,$+$\,CO\,$\rightarrow$\,H\,$+$\,CO$_2$, studies by Schatz and co-workers and by Guo and co-workers have examined the influence of the entrance-channel well on thermal rates and the different effects of OH and CO vibrational excitation on reactivity \cite{Medvedev2004,Li2012HOCO}. In \emph{atmospheric chemistry}, barrierless O$_2$ capture by organic radicals initiates and propagates the peroxy-radical chemistry that controls OH recycling and secondary organic aerosol formation in isoprene oxidation \cite{Peeters2009,Peeters2014,Wennberg2018,Muller2019,Szabo2024hpald,Berndt2019,Novelli2020,Teng2017}, with capture rate coefficients and site-specific branching ratios accessible to on-the-fly QCT \cite{Szabo2025capture,Miyoshi1990,Fernandes2006}. In \emph{astrochemical} environments, classical trajectories give access to radiative association \cite{SzaboGustafsson2023} and to collision-induced absorption spectra of atomic and molecular gas mixtures \cite{Fakhardji2019,Fakhardji2020,Fakhardji2021}, processes that shape the radiative properties and chemistry of planetary and stellar atmospheres. Finally, \emph{gas--surface} collisions---from energy transfer of rare gases and small molecules at self-assembled monolayers \cite{BosioHase1997,MartinezNunez2007} through hyperthermal erosion chemistry \cite{MintonGarton2001,Yan2004} to the collision-induced and surface-induced dissociation of biomolecular ions \cite{Meroueh2002,Barnes2011}---constitute a fourth broad application area, in which trajectory simulations have been instrumental in connecting measured energy disposal and fragmentation patterns to atomistic mechanisms.

Not all important collisions are reactive. In \emph{elastic} collisions only the direction of the relative motion changes; in \emph{inelastic} collisions energy is transferred between relative translation and the internal (vibrational, rotational) degrees of freedom of the partners, or between the internal modes themselves (V--T, V--R, V--V transfer). Competition between reaction and vibrational relaxation has also been studied for H atoms colliding with highly vibrationally excited H$_2$O \cite{Barnes2001}. Full-dimensional QCT studies have also characterized energy transfer in H$_2$CO--H$_2$ collisions \cite{Guo2021} and vibration--vibration exchange between excited CO molecules \cite{Chen2020}. In CO--CO collisions, combined experimental, QCT, and quantum-scattering studies revealed a ``molecular square-dancing'' mechanism in which substantial rotational excitation of both partners accompanies little angular deflection \cite{Sun2020}. Collisional energy transfer of highly vibrationally excited molecules is the elementary step that controls the pressure dependence of essentially every unimolecular and recombination reaction: the energy-transfer kernel $P(E,E')$ and its moments, most prominently the average energy transferred per collision $\langle \Delta E \rangle$, are required inputs of master-equation models of thermal kinetics \cite{Lendvay2019,OrefTardy1990,Barker2001,Troe1977}. Because state-resolved experiments on highly excited polyatomics are extraordinarily difficult, classical trajectory calculations have become the principal source of detailed, system-specific energy-transfer data---cross sections, probability distributions, moments, and mechanistic information such as gateway modes and the role of transient complex formation \cite{Lendvay2019}. A general-purpose QCT code must therefore treat reactive and nonreactive trajectories on the same footing and provide the corresponding analysis machinery.

Beyond reactive encounters, QCT can also describe inelastic collisions relevant to \emph{astrochemistry} \cite{Loreau2018COH2O}. In astrophysical environments, molecular level populations often depart from local thermodynamic equilibrium. State-resolved collisional rate coefficients enter the excitation models used to infer gas temperatures, densities, and molecular abundances from observed spectra. QCT calculations have provided rotational (de)excitation rates for H$_2$O--H$_2$ collisions over extensive sets of levels \cite{Faure2007}, as well as vibrational-relaxation rates for the water bending mode on a full-dimensional H$_2$O--H$_2$ PES \cite{Faure2005}. For larger molecules, QCT has also complemented coupled-channel calculations of HC$_3$N rotational excitation by para- and ortho-H$_2$, providing collisional data for astronomical excitation models \cite{Faure2016}. For CO--H$_2$O, relevant to environments in which water is an important collider, Loreau, Faure, and Lique compared QCT, coupled-states, and statistical adiabatic-channel calculations with close-coupling benchmarks on a five-dimensional rigid-rotor PES \cite{Loreau2018COH2O}. Their results illustrate complementary regimes: the statistical treatment is effective at low collision energies for this complex-forming system, whereas QCT becomes more advantageous at higher energies. Trajectory methods thus extend the accessible range of molecular size, excitation, and temperature, while low-temperature state-to-state predictions require particular care because of quantum thresholds and resonances.

\subsection{The quasiclassical trajectory method and this work}

The QCT method \cite{TruhlarMuckerman1979,Peslherbe1999,Hase1998encyc} occupies a pragmatic middle ground between full quantum scattering---exact but currently limited to few-atom systems---and purely classical molecular dynamics. Initial conditions are drawn from ensembles of classical states that correspond to preselected quantum-mechanical vibrational and rotational states of the reactants (``quasiclassical'' sampling), the trajectories are then propagated with classical mechanics, and product states are assigned by semiclassical binning of the final actions, either by standard histogram binning or by Gaussian-weighted binning \cite{BonnetRayez1997,BonnetRayez2004,Banares2003}. Comparisons with experiment and quantum dynamics have demonstrated useful accuracy for many direct bimolecular reactions, short-time unimolecular decompositions, and fragmentations on repulsive surfaces, particularly where tunneling, interference, and zero-point-energy leakage are not dominant \cite{Paranjothy2013,HaseBuchowski1982,Schatz1983,Miller1989}; known limitations, chiefly zero-point-energy leakage in long-lived complexes and the absence of tunneling, are well documented and can be diagnosed and partially mitigated \cite{Miller1989,Guo2012}. The great strengths of the method are its generality (any number of atoms, any PES), its ability to deliver semiclassically state-resolved cross sections and rate coefficients, and the direct mechanistic insight afforded by trajectory animation and event counting.

In this paper we describe \Smite, a Python toolkit that brings molecular preparation, trajectory propagation, stationary-point characterization, and trajectory observables into one framework. The internal preparation is explicitly mode resolved, while the trajectory engine can use direct electronic-structure evaluations or analytical and machine-learned PESs. This separation makes it possible to compare preparation protocols and potential models without replacing the dynamics workflow. Photoionization preparation and a PES-based surface-hopping module extend the range of accessible launch conditions and electronic-state dynamics. Section~\ref{sec:qct} defines the implemented initial-condition distributions, with particular attention to vibrations, rotations, and their coupling. Section~\ref{sec:code} describes the architecture and additional modules. Section~\ref{sec:analysis} connects the recorded trajectories to dynamical and spectroscopic observables, distinguishing implemented per-trajectory analysis from ensemble estimators and optional external post-processing. Section~\ref{sec:benchmarks} describes the validation tests and example systems that accompany the code.

\section{Quasiclassical trajectory methodology}
\label{sec:qct}

\subsection{Preparation of a molecular phase-space ensemble}
\label{sec:ensemble}

A trajectory is specified by a Cartesian phase point $(\mathbf q,\mathbf p)$, atom identities and masses, and an electronic-state or PES model. In QCT calculations, quantization enters primarily through the preparation and interpretation of that phase point; subsequent nuclear propagation is classical \cite{TruhlarMuckerman1979,Peslherbe1999}. The preparation procedure should therefore identify the distribution being sampled, rather than only its nominal temperature or total energy. \Smite\ separates vibrational sampling, rotational preparation, overall orientation, and, for collisions, relative translation. These choices can be combined to construct equilibrium-like ensembles or deliberately nonequilibrium mode-selective ensembles.

Table~\ref{tab:sampling} summarizes the implemented prescriptions. A fixed set of harmonic mode energies is a phase ensemble on a product of oscillator orbits. It is not, by itself, an ergodic microcanonical ensemble of the full anharmonic molecular Hamiltonian. Likewise, sampling thermal harmonic quantum populations and classical polyatomic rotations defines a useful mixed quantum--classical preparation, not an exact quantum thermal density operator. These distinctions are particularly relevant for flexible molecules and for excitation close to dissociation.

\begin{table*}[t]
\centering
\caption{\label{tab:sampling}Initial-condition choices implemented in \Smite.
Energies and distributions refer to the preparation model; the subsequent
trajectory can use a different, anharmonic PES.}
\small
\renewcommand{\arraystretch}{1.18}
\setlength{\tabcolsep}{5pt}

\begin{tabular}{@{}llll@{}}
\hline
\noalign{\smallskip}
\tcell{2.5cm}{Degree of freedom} & \tcell{3.0cm}{Prescription} & \tcell{\dimexpr\textwidth-8.9cm-6\tabcolsep\relax}{Sampled quantity} & \tcell{3.4cm}{Scope} \\
\noalign{\smallskip}
\hline
\noalign{\smallskip}
\tcell{2.5cm}{Harmonic vibration} & \tcell{3.0cm}{Fixed quantum number (Q)} & \tcell{\dimexpr\textwidth-8.9cm-6\tabcolsep\relax}{$E_k=\hbar\omega_k(n_k+1/2)$; uniform phase} & \tcell{3.4cm}{Harmonic diatoms and polyatomics} \\
\tcell{2.5cm}{} & \tcell{3.0cm}{Fixed energy (E)} & \tcell{\dimexpr\textwidth-8.9cm-6\tabcolsep\relax}{Prescribed $E_k$; uniform phase} & \tcell{3.4cm}{Mode energy need not be quantized} \\
\tcell{2.5cm}{} & \tcell{3.0cm}{Thermal population (T)} & \tcell{\dimexpr\textwidth-8.9cm-6\tabcolsep\relax}{Boltzmann-distributed integer $n_k$; uniform phase} & \tcell{3.4cm}{Physical mode frequency is used} \\
\tcell{2.5cm}{} & \tcell{3.0cm}{Ground-state Wigner (W)} & \tcell{\dimexpr\textwidth-8.9cm-6\tabcolsep\relax}{Gaussian $Q_k$ and $P_k$} & \tcell{3.4cm}{Distributed energies; ground state only} \\
\tcell{2.5cm}{Polyatomic rotation} & \tcell{3.0cm}{Fixed $J$ or thermal} & \tcell{\dimexpr\textwidth-8.9cm-6\tabcolsep\relax}{Fixed $|\mathbf J|$, or Gaussian principal-axis components} & \tcell{3.4cm}{General rigid top; linear-axis rotation excluded} \\
\tcell{2.5cm}{Diatomic rotation} & \tcell{3.0cm}{Fixed $J$ or thermal} & \tcell{\dimexpr\textwidth-8.9cm-6\tabcolsep\relax}{Discrete $J$, weighted by $(2J+1)$ in a thermal draw} & \tcell{3.4cm}{Rigid-rotor preparation model} \\
\tcell{2.5cm}{Morse diatom} & \tcell{3.0cm}{Selected or thermal $(n,J)$} & \tcell{\dimexpr\textwidth-8.9cm-6\tabcolsep\relax}{Coupled bound rovibrational state and radial phase} & \tcell{3.4cm}{Fixed energy denotes total rovibrational energy} \\
\tcell{2.5cm}{Saved MD} & \tcell{3.0cm}{Phase-point reuse} & \tcell{\dimexpr\textwidth-8.9cm-6\tabcolsep\relax}{A stored coordinate--momentum pair} & \tcell{3.4cm}{Ensemble inherited from the source run} \\
\tcell{2.5cm}{Relative translation} & \tcell{3.0cm}{Fixed or thermal} & \tcell{\dimexpr\textwidth-8.9cm-6\tabcolsep\relax}{$\Ecoll$, impact parameter, and orientation} & \tcell{3.4cm}{Thermal collision energies are flux weighted} \\
\tcell{2.5cm}{Photoionization} & \tcell{3.0cm}{Levels 0--3} & \tcell{\dimexpr\textwidth-8.9cm-6\tabcolsep\relax}{Neutral-to-ion momentum mapping} & \tcell{3.4cm}{Energy constraints and optional recoil} \\
\noalign{\smallskip}
\hline
\end{tabular}
\end{table*}

\subsection{Normal-mode coordinates and conventions}
\label{sec:semicl}

For a nonrigid polyatomic fragment, the reference geometry $\mathbf q_{\mathrm{eq}}$ and Cartesian Hessian $\mathbf H$ define a harmonic preparation model. Let $\mathbf M$ be the diagonal $3N\times3N$ mass matrix, with each atomic mass repeated three times. Let $\mathbf\Pi$ project out the mass-weighted translation and rotation vectors in an Eckart construction \cite{Eckart1935}. The projected Hessian and its vibrational eigenvectors satisfy
\begin{align}
 \mathbf H_{\mathrm{vib}}&=\mathbf\Pi\mathbf M^{-1/2}\mathbf H\mathbf M^{-1/2}\mathbf\Pi,\nonumber\\
 \mathbf H_{\mathrm{vib}}\mathbf U&=\mathbf U\bm\Omega^2,\qquad \mathbf U^T\mathbf U=\mathbf 1,
 \label{eq:masshessian}
\end{align}
where $\bm\Omega=\operatorname{diag}(\omega_1,\ldots,\omega_f)$ and $f=3N-6$ for a nonlinear molecule or $3N-5$ for a linear molecule. The matrix stored by the normal-mode machinery is the Cartesian displacement matrix
\begin{equation}
 \mathbf L=\mathbf M^{-1/2}\mathbf U,\qquad
 \mathbf L^T\mathbf M\mathbf L=\mathbf 1.
 \label{eq:Lconvention}
\end{equation}
It is essential to distinguish $\mathbf L$ from the orthonormal mass-weighted eigenvectors $\mathbf U$: adding another factor of $\mathbf M^{-1/2}$ to $\mathbf L$ would apply mass weighting twice.

With mass-weighted mode coordinates $\mathbf Q$ and conjugate momenta $\mathbf P=\dot{\mathbf Q}$, the Cartesian transformation is
\begin{align}
 \mathbf q&=\mathbf q_{\mathrm{eq}}+\mathbf L\mathbf Q,
 \label{eq:trafoq}\\
 \mathbf p_{\mathrm{vib}}&=\mathbf M\mathbf L\mathbf P.
 \label{eq:trafop}
\end{align}
The inverse projection, in the same molecular frame, is
\begin{equation}
 \mathbf Q=\mathbf L^T\mathbf M(\mathbf q-\mathbf q_{\mathrm{eq}}),
 \qquad \mathbf P=\mathbf L^T\mathbf p.
 \label{eq:backproject}
\end{equation}
These definitions give the harmonic energy $H_{\mathrm{vib}}=\sum_k(P_k^2+\omega_k^2Q_k^2)/2$. If a molecule has subsequently been rotated in the laboratory frame, its coordinates and momenta must first be transformed consistently back to the reference frame before applying Eq.~\eqref{eq:backproject}.

\subsection{Fixed-action, fixed-energy, and thermal vibrational sampling}
\label{sec:viboptions}

For a selected harmonic quantum number $n_k$, the mode energy is
\begin{equation}
 E_k=\hbar\omega_k(n_k+\tfrac12).
 \label{eq:Ek}
\end{equation}
Alternatively, a positive or zero $E_k$ can be supplied directly. Both choices use one uniformly distributed phase $\phi_k=2\pi\xi_k$, with $\xi_k\in[0,1)$, to sample
\begin{align}
 Q_k&=\frac{\sqrt{2E_k}}{\omega_k}\cos\phi_k,
 \label{eq:Qk0}\\
 P_k&=-\sqrt{2E_k}\sin\phi_k.
 \label{eq:Qdotk0}
\end{align}
Positions and momenta share the same phase, so that every draw has exactly the prescribed harmonic mode energy. Independent position and momentum phases would destroy this constraint. Fixed-energy sampling reports $n_k^{\mathrm{eff}}=E_k/(\hbar\omega_k)-1/2$ as a diagnostic; this continuous quantity is not an assigned quantum eigenstate.

For thermal mode preparation at $T_k$, the quantum number is sampled from
\begin{equation}
 p_k(n)=(1-e^{-\beta_k\hbar\omega_k})e^{-\beta_k\hbar\omega_k n},
 \quad \beta_k=(\kb T_k)^{-1},
\end{equation}
using inverse transformation,
\begin{equation}
 n_k=\left\lfloor-\frac{\kb T_k}{\hbar\omega_k}\ln(1-\xi)\right\rfloor.
 \label{eq:thermal}
\end{equation}
A fresh uniform phase then supplies $(Q_k,P_k)$. The mean harmonic energy is
\begin{equation}
 \langle E_k\rangle=\frac{\hbar\omega_k}{2}
 \coth\!\left(\frac{\hbar\omega_k}{2\kb T_k}\right).
 \label{eq:thermalmean}
\end{equation}
Zero-temperature preparation is obtained by selecting $n_k=0$. Global defaults can be overridden for individual modes through \code{fix\_quantum}, \code{fix\_energy}, and \code{fix\_temp}. Thus, a selected normal-mode excitation can be embedded in an otherwise thermal or zero-point ensemble. Different mode temperatures define a nonequilibrium preparation and should be reported as such.

The current implementation uses each mode's physical frequency in both its thermal population and its phase-space mapping. It does not impose a low-frequency population floor. This preserves the chosen harmonic population law, but it cannot make a harmonic treatment of a hindered rotation or a large-amplitude torsion physically adequate. A saved anharmonic MD ensemble can be preferable for those motions. Imaginary and zero-frequency modes are excluded by the default sampling policy; a legacy absolute-frequency option is explicitly flagged because it does not describe equilibrium sampling of an unstable mode.

\subsection{Ground-state Wigner sampling and preparation diagnostics}
\label{sec:wigner}

For harmonic ground-state Wigner sampling \cite{Wigner1932}, the distribution of one mode is
\begin{equation}
 W_0(Q_k,P_k)=\frac{1}{\pi\hbar}
 \exp\!\left[-\frac{\omega_kQ_k^2}{\hbar}
             -\frac{P_k^2}{\hbar\omega_k}\right].
 \label{eq:wigner}
\end{equation}
Hence $Q_k$ and $P_k$ are independent zero-mean Gaussian variables with
\begin{equation}
 \langle Q_k^2\rangle=\frac{\hbar}{2\omega_k},\qquad
 \langle P_k^2\rangle=\frac{\hbar\omega_k}{2}.
\end{equation}
The implementation uses an equivalent polar sampling of the two Gaussian variables. Unlike fixed-$n_k=0$ QCT preparation, a Wigner ensemble does not constrain the classical harmonic energy of each sample to $\hbar\omega_k/2$: that value is its ensemble mean. Individual samples can lie below or above it. The two prescriptions should therefore be compared through their distributions and intended application, rather than treated as interchangeable zero-point preparations. Wigner initialization also does not make subsequent classical propagation exact quantum dynamics.

The global \code{Wigner} option and per-mode \code{fix\_wigner} overrides currently describe the harmonic ground state. Finite-temperature Wigner sampling and excited-state Wigner distributions are not implemented by this option. In particular, thermal quantum-number sampling followed by fixed-energy phase sampling is a different distribution from the finite-temperature Wigner function, although both reproduce Eq.~\eqref{eq:thermalmean}.

After Cartesian assembly, optional diagnostics back-project the phase point with Eq.~\eqref{eq:backproject} and report
\begin{equation}
 E_k^{\mathrm{rec}}=\tfrac12P_k^2+\tfrac12\omega_k^2Q_k^2,
 \qquad \Delta E_k=E_k^{\mathrm{rec}}-E_k^{\mathrm{requested}}.
\end{equation}
For a Wigner draw, the recorded reference energy is the ensemble ground-state mean, so a nonzero single-sample difference is expected. The diagnostic is applied after vibrational preparation, before the subsequent rotational adjustment and random orientation. Those later operations, and evaluation on an anharmonic PES, must be considered separately when assessing the final initial-condition energy. In particular, fixed harmonic actions do not guarantee an identical total energy on the full PES for every sampled geometry. The code also detects atomic overlaps; such rejection changes the sampled distribution and should be monitored if frequent.

\subsection{Rotational preparation and angular-momentum control}
\label{sec:rot}

Rotations are prepared in the fragment center-of-mass frame. Let $\mathbf r_a$ denote atom $a$'s position relative to the center of mass, and define
\begin{align}
 \mathbf I&=\sum_a m_a\big[(\mathbf r_a\cdot\mathbf r_a)\mathbf 1
                         -\mathbf r_a\mathbf r_a^T\big],\\
 \mathbf J_{\mathrm{vib}}&=\sum_a\mathbf r_a\times\mathbf p_{a,\mathrm{vib}}.
\end{align}
Finite normal-mode displacements can carry angular momentum even when the infinitesimal modes were projected at equilibrium. To realize a desired total $\mathbf J_{\mathrm{target}}$, the code applies
\begin{align}
 \bm\omega_{\mathrm{corr}}&=\mathbf I^{+}
          (\mathbf J_{\mathrm{target}}-\mathbf J_{\mathrm{vib}}),\\
 \mathbf p_a&\leftarrow\mathbf p_{a,\mathrm{vib}}
             +m_a\bm\omega_{\mathrm{corr}}\times\mathbf r_a,
 \label{eq:rotcorrection}
\end{align}
where $\mathbf I^+$ is the inverse or, for a linear rotor, its pseudoinverse. The realized angular momentum is checked against the target. This correction enforces angular momentum, but can alter the mode-resolved kinetic energies; exact fixed harmonic actions and exact finite-displacement angular-momentum constraints should not be assumed to survive simultaneously without a separate check.

With \code{Jfix}, the requested semiclassical magnitude is
\begin{equation}
 |\mathbf J_{\mathrm{target}}|=\hbar\sqrt{J(J+1)}.
 \label{eq:Jfix}
\end{equation}
For nonlinear fragments its direction is uniform on the sphere. For linear fragments, including diatomics, it lies on the circle perpendicular to the molecular axis, because rotation about an axis with zero moment of inertia cannot be represented by Cartesian rigid-body motion. For an asymmetric top, fixing $J$ does not specify a complete quantum rotational eigenstate: the rotational energy also depends on direction relative to the principal axes, and no independent $K$-state selection is implied.

Thermal polyatomic rotation is sampled from the classical canonical distribution of a general rigid top. If $I_\alpha$ are the nonzero principal moments, its principal-axis angular momenta obey
\begin{equation}
 J_\alpha=\sqrt{I_\alpha\kb T_{\mathrm{rot}}}\,z_\alpha,
 \qquad z_\alpha\sim\mathcal N(0,1),
 \label{eq:thermalrot}
\end{equation}
independently. Thus
\begin{equation}
 p(\mathbf J)\propto
 \exp\!\left[-\frac{1}{2\kb T_{\mathrm{rot}}}
                  \sum_{\alpha:I_\alpha>0}\frac{J_\alpha^2}{I_\alpha}\right].
\end{equation}
The sampled vector is rotated back to the laboratory frame before Eq.~\eqref{eq:rotcorrection} is applied. The mean rotational energy is $3\kb T_{\mathrm{rot}}/2$ for a nonlinear rotor and $\kb T_{\mathrm{rot}}$ for a linear rotor. This construction treats asymmetric, symmetric, and spherical tops in one expression; no prolate/oblate approximation is required. It is a classical rotational distribution conditional on the geometry being sampled.

For harmonic diatomic preparation, thermal rotation instead draws a discrete $J$ from
\begin{equation}
 p(J)=\frac{(2J+1)e^{-\beta B J(J+1)}}{Z_{\mathrm{rot}}},
 \quad B=\frac{\hbar^2}{2I},\quad I=\mu r^2,
 \label{eq:diatomrot}
\end{equation}
using the sampled bond length $r$ and reduced mass $\mu$. The discrete sum is extended until its remaining tail is negligible. The default degeneracy factor does not include species-specific nuclear-spin statistical weights; ortho/para selection requires a correspondingly restricted preparation.

\subsection{Harmonic and rotating-Morse diatomic fragments}
\label{sec:diatom}

Harmonic diatoms support the same Q, E, T, and ground-state W prescriptions as a polyatomic normal mode. In a fixed-energy draw,
\begin{equation}
 r-r_e=\sqrt{\frac{2E_v}{\mu\omega^2}}\cos\phi,\qquad
 p_r=-\sqrt{2\mu E_v}\sin\phi,
\end{equation}
after which the two-atom phase point is constructed at zero center-of-mass momentum and rotational momentum is added.

For anharmonic preparation, a rotating Morse model uses $r_e$, the well depth $D_e$, and range parameter $a$, with $V_M(r)=D_e[1-e^{-a(r-r_e)}]^2$ and $\omega_0=a\sqrt{2D_e/\mu}$. Following Porter, Raff, and Miller \cite{PorterRaffMiller1975}, the implemented rovibrational energy is
\begin{align}
 E_{nJ}={}&E_h+E_R-\frac{E_h^2}{4D_e}
             -\frac{E_R^2}{D_ea^2r_e^2}\nonumber\\
          &-\frac{3[1-(ar_e)^{-1}]E_RE_h}{2ar_eD_e},
 \label{eq:morseenergy}\\
 E_h={}&\hbar\omega_0(n+\tfrac12),\qquad
 E_R=\frac{\hbar^2J(J+1)}{2\mu r_e^2}.
\end{align}
The coordinate and radial momentum are sampled together from the corresponding rotating-Morse construction. Equation~\eqref{eq:morseenergy} is the preparation model, including centrifugal and rotation--vibration corrections; it is not an exact quantum solution of an arbitrary diatomic PES.

Thermal Morse preparation enumerates the bound states supported by this model and samples
\begin{equation}
 p(n,J)=\frac{(2J+1)e^{-\beta E_{nJ}}}{Z_{\mathrm{bound}}}.
 \label{eq:morsecoupled}
\end{equation}
When either $n$ or $J$ is fixed, the distribution is conditional on that value. Simultaneously thermal vibration and rotation use a common temperature. A fixed Morse energy denotes total bound rovibrational energy relative to the potential minimum and therefore requires fixed $J$; the lower physical branch of Eq.~\eqref{eq:morseenergy} is inverted to obtain a continuous action. Unsupported unbound states are rejected. Ground-state Wigner sampling is available for the harmonic model, not for the Morse model. The preparation model and the propagation PES should be matched or their energy differences quantified.

\subsection{Saved MD ensembles, orientation, and reproducibility}
\label{sec:mdsampling}

The \code{fromMD} route draws a stored coordinate--momentum pair from a user-selected part of a previous MD trajectory. It can retain anharmonic and conformational information that is absent from a single-reference harmonic model. The resulting distribution inherits the source dynamics, thermostat, equilibration interval, and time correlation; random frame selection alone does not guarantee independent canonical samples. In the fragment workflow, subsequent center-of-mass removal and any requested rotational reassignment also alter that source ensemble. To retain its internal rotational preparation, rotational resampling should be disabled explicitly. The separate \code{MDprimitive} route assigns Maxwell--Boltzmann momenta at a given geometry and requires subsequent equilibration if a configurational thermal ensemble is intended.

For isotropic gas-phase ensembles, the same random rotation is applied to positions and momenta. Euler angles are drawn uniformly in the two azimuths and uniformly in the cosine of the polar angle, giving the uniform rotational measure. Independently sampled reactants then acquire their relative translation as described below. Seeding initializes the complete preparation stream, and parallel trajectory seeds are derived from a base seed and trajectory index, so the intended sequence is independent of worker scheduling. The seed, preparation choices, electronic model, and geometry reference should accompany reported observables.

\subsection{Bimolecular relative initial conditions}
\label{sec:bimol}

Two prepared fragments are combined by assigning their initial separation $R_{\mathrm{ini}}$, impact parameter $b$, and relative translational energy $\Ecoll$. At fixed energy, the relative speed is $v_{\mathrm{rel}}=\sqrt{2\Ecoll/\mu}$, where $\mu=M_AM_B/(M_A+M_B)$. With the incoming relative velocity along $x$, fragment B is displaced by $\sqrt{R_{\mathrm{ini}}^2-b^2}$ along $x$ and by $b$ along $z$. The velocities
\begin{equation}
 v_A=\frac{M_B}{M_A+M_B}v_{\mathrm{rel}},\qquad
 v_B=-\frac{M_A}{M_A+M_B}v_{\mathrm{rel}}
\end{equation}
give zero total center-of-mass momentum. $R_{\mathrm{ini}}$ must be large enough that the chosen preparation represents separated reactants; long-range interactions may require a separation-convergence study.

For direct thermal rate sampling, the code draws collision energies from the \emph{flux-weighted} Maxwell distribution,
\begin{equation}
 p_{\mathrm{flux}}(E)=\frac{E}{(\kb T)^2}e^{-E/(\kb T)},\qquad
 \Ecoll=-\kb T\ln(\xi_1\xi_2).
 \label{eq:ecolltherm}
\end{equation}
This differs from the unweighted relative-energy distribution, which is proportional to $\sqrt E\,e^{-E/(\kb T)}$. The extra factor of relative speed in a rate coefficient explains the flux weighting and the estimator in Sec.~\ref{sec:rates}.

The impact parameter is either fixed, sampled uniformly in $b$, or sampled uniformly over the collision disk:
\begin{equation}
 b=\bmax,\qquad b=\bmax\xi,\qquad
 b=\bmax\sqrt\xi,
 \label{eq:bsampling}
\end{equation}
respectively. These choices have different statistical weights. A fixed-$b$ calculation estimates an opacity at that $b$; it does not directly give an integrated cross section. Linear sampling emphasizes small impact parameters and requires the weighting given in Sec.~\ref{sec:rates}. The two fragments can have independently selected internal temperatures or vibrational states, so a thermal collision-energy draw alone does not imply a fully thermal rate coefficient.

Reaction termination can use a global separation threshold or named channels defined by conjunctions of pair-distance conditions. In collisions, \code{Rstop} refers to the separation of the original reactant centers of mass; in unimolecular propagation it tests the largest atom-pair separation. Named channels are checked at every step, independently of trajectory output frequency. An optional persistence count requires the same channel to hold over successive steps. Distance-based termination is an operational classifier and should be tested for sensitivity to thresholds, recrossing, and the maximum propagation time.

\subsection{Molecule--surface collisions}
\label{sec:surface}

A finite atomic fragment can be designated as a surface and oriented with its normal along the laboratory $x$ direction. The projectile is prepared independently. A fixed incidence angle can be supplied; otherwise the cosine of the angle is sampled uniformly over a spherical cap,
\begin{equation}
 \cos\theta=1-\xi(1-\cos\theta_{\max}),
 \label{eq:skew}
\end{equation}
with randomized azimuth. This is uniform solid-angle sampling, not the cosine-weighted angular flux of an equilibrium gas impinging on a plane; the latter would require appropriate sampling or reweighting. The placement preserves the requested initial separation and perpendicular impact parameter. Surface phase points can be reused from previous MD preparation \cite{BosioHase1997,MartinezNunez2007,Meroueh2002}. The present setup describes finite molecular or cluster surface models and does not itself imply periodic boundaries, a bulk substrate, or an implicit heat bath.

\section{Program structure and capabilities}
\label{sec:code}

\subsection{Architecture, simulation objects, and units}

\begin{figure*}[t]
\centering
\includegraphics{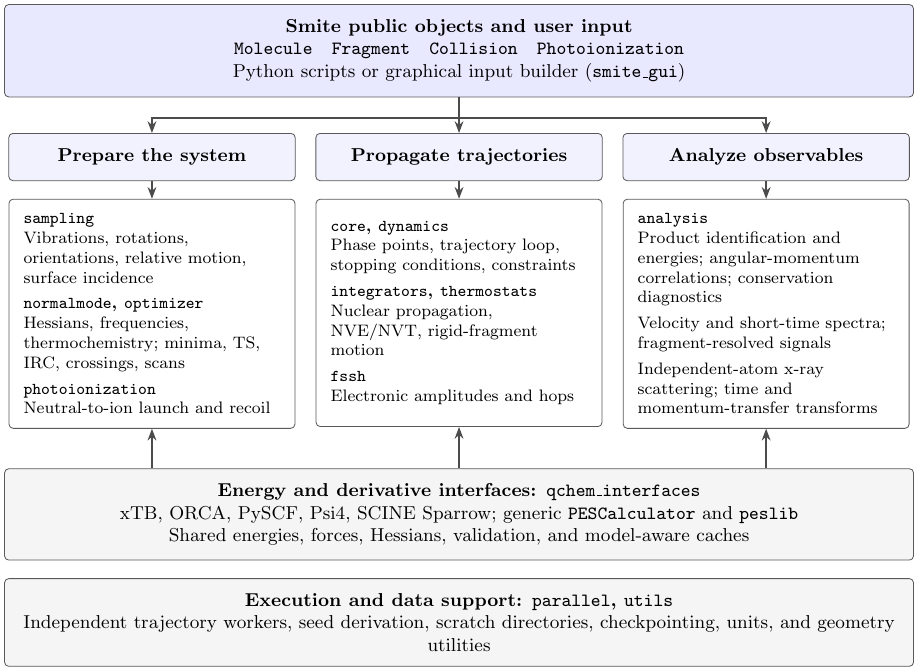}
\caption{\label{fig:architecture}Module map of \Smite. The upper branches organize capabilities by their role in a calculation; the lower bands provide shared services. This is a functional map, not a Python inheritance or import graph. The present FSSH state adapter uses PES calculators, whereas single-surface nuclear dynamics and stationary-point tools also use the electronic-structure backends. The input builder generates scripts for a subset of the Python interface.}
\end{figure*}

Figure~\ref{fig:architecture} summarizes the modules and their roles in a calculation. The public import exposes \code{Molecule}, \code{Fragment}, \code{Collision}, and \code{Photoionization}. \code{Molecule} holds atoms, masses, Cartesian positions and momenta, the electronic model, and trajectory state. \code{Fragment} specializes preparation for atoms, diatoms, and polyatomics. \code{Collision} combines two fragments and adds relative-motion sampling and collision analysis. \code{Photoionization} prepares an ionic phase point from a neutral source and then uses the common molecular propagator. Stationary-point tools can also be called independently of a trajectory.

Internal propagation uses atomic units: coordinates in bohr, momenta in atomic units, energies in hartree, and time in atomic units. The principal user interfaces accept XYZ geometries and collision distances in \AA, nuclear and electronic trajectory timesteps in femtoseconds, collision energies in kJ\,mol$^{-1}$, and temperatures in kelvin. Photoionization photon, electron, and binding energies are supplied in eV. API-specific inputs, particularly directly prescribed mode energies, should follow their documented units. Explicit conversion at interfaces avoids mixing the propagation convention with spectroscopy and experimental energy conventions.

\subsection{Nuclear propagation, stopping conditions, and thermostats}
\label{sec:prop}

On one electronic surface, the nuclear equations are
\begin{equation}
 \dot{\mathbf q}=\mathbf M^{-1}\mathbf p,\qquad
 \dot{\mathbf p}=-\nabla V(\mathbf q).
\end{equation}
Available integrators include velocity Verlet and leapfrog, St\"ormer--Verlet, fourth-order Runge--Kutta, higher-order symplectic compositions, symplectic partitioned Runge--Kutta schemes, and Adams--Bashforth--Moulton predictor--corrector propagation \cite{Verlet1967,Swope1982,Yoshida1990,Hairer2006,HairerNorsettWanner1993}. Time-step convergence and energy conservation must be established for the selected PES and initial energy; a high formal integration order does not compensate for noisy forces or unconverged electronic calculations.

The trajectory loop inspects termination conditions at every nuclear step, including the initial and final states, and writes a terminal frame independently of the normal output interval. For a continuing trajectory it applies the nuclear integrator, an optional thermostat, and an optional electronic propagation step. Energy drift, trajectory temperature, termination reason, channel, and final step are recorded. These diagnostics permit subsequent trajectory selection; the default dynamics driver does not implement a universal automatic energy-drift rejection policy.

\begin{table}[t]
\caption{\label{tab:thermostats}Thermalization options. An isolated production trajectory uses no thermostat.}
\small
\renewcommand{\arraystretch}{1.18}
\setlength{\tabcolsep}{4pt}
\begin{tabular}{@{}ll@{}}
\toprule
\tcell{1.75cm}{Thermostat} & \tcell{\dimexpr\columnwidth-1.75cm-2\tabcolsep\relax}{Role and interpretation} \\
\midrule
\tcell{1.75cm}{Berendsen} & \tcell{\dimexpr\columnwidth-1.75cm-2\tabcolsep\relax}{Weak-coupling temperature relaxation; not an exact canonical sampler \cite{Berendsen1984}} \\
\tcell{1.75cm}{Andersen} & \tcell{\dimexpr\columnwidth-1.75cm-2\tabcolsep\relax}{Stochastic Maxwell momentum updates \cite{Andersen1980}} \\
\tcell{1.75cm}{Nos\'e--Hoover} & \tcell{\dimexpr\columnwidth-1.75cm-2\tabcolsep\relax}{Extended-variable dynamics; canonical sampling requires adequate ergodicity \cite{Nose1984,Hoover1985}} \\
\tcell{1.75cm}{GLE} & \tcell{\dimexpr\columnwidth-1.75cm-2\tabcolsep\relax}{Colored-noise dynamics with auxiliary variables and user-defined parameters \cite{Ceriotti2009,Ceriotti2010}} \\
\bottomrule
\end{tabular}
\end{table}

Table~\ref{tab:thermostats} lists the thermostat families. In the Andersen interface, the supplied collision time $\tau$ is a mean waiting time and the probability of a refresh during $\Delta t$ is $1-e^{-\Delta t/\tau}$. Center-of-mass removal and rigid constraints are treated as physical projections; they are not represented by freezing an arbitrary suffix of Cartesian components. The kinetic temperature is $T=2K/(N_{\mathrm{dof}}\kb)$, with the degrees of freedom adjusted for the specified projections and constraints. For an isolated collision, a thermostat would alter the energy exchange being measured and is normally used only to prepare the starting ensemble.

\subsection{Rigid fragments and constrained dynamics}
\label{sec:rigid}

Selected, nonoverlapping atom groups can be constrained to remain rigid. The implementation combines distance constraints with momentum projection and reversible rigid-body drift; the constrained nuclear path is currently exposed with Verlet or leapfrog selection. For a constrained pair,
\begin{align}
 g_{ab}(\mathbf q)&=|\mathbf r_a-\mathbf r_b|^2-d_{ab}^2=0,\\
 (\mathbf r_a-\mathbf r_b)\cdot
 \left(\frac{\mathbf p_a}{m_a}-\frac{\mathbf p_b}{m_b}\right)&=0.
\end{align}
SHAKE/RATTLE-style position and momentum corrections enforce both conditions to prescribed tolerances \cite{Ryckaert1977,Andersen1983}. A rigid nonlinear group retains six external degrees of freedom, a rigid linear group five, and a rigid diatom removes one bond-stretch coordinate. The constraints retain molecular rotation and translation while eliminating internal deformation. They are useful for separating orientational effects from intramolecular energy transfer, but the resulting rigid-fragment model must be identified explicitly when interpreting reaction or relaxation observables.

Rigid-fragment propagation can also be applied to rotationally inelastic collisions on reduced-dimensionality PESs that retain the intermolecular separation and relative orientations while fixing the intramolecular coordinates, as in the five-dimensional CO--H$_2$O treatment of Ref.~\cite{Loreau2018COH2O}. In \Smite, a user-supplied calculator must map the reduced-coordinate potential and its derivatives consistently to the Cartesian energy/force interface. This permits rotational energy transfer while excluding vibrational excitation of the constrained fragments; a flexible PES and unconstrained internal coordinates are needed to describe the latter.

\subsection{Electronic-structure backends and PES calculators}
\label{sec:backends}
\label{sec:pesif}

All single-surface energy and force calls use a shared input dictionary selecting the backend, method settings, charge, spin multiplicity, thread count, and scratch location. The implemented choices are external xTB \cite{Bannwarth2019,Grimme2021xtb}, ORCA \cite{Neese2012,Neese2025}, in-process PySCF \cite{Sun2018pyscf} and Psi4 \cite{Smith2020psi4}, SCINE Sparrow through binary or Python interfaces \cite{Husch2018}, and the generic PES interface. Backend-specific capabilities remain those of the implemented adapter and selected electronic method; an interface does not imply support for every method available in its parent quantum-chemistry package.

For direct Born--Oppenheimer dynamics, the adapter obtains an energy and gradient at the requested geometry, with the common force convention $\mathbf F=-\nabla V$. Selected backends reuse orbital information between nearby geometries. Energy/force caching checks coordinates, atom ordering, and the electronic-model fingerprint, so changing a charge, multiplicity, method, or surface invalidates the old result. Scratch isolation, informative backend errors, and optional cleanup-and-retry support long trajectory campaigns. The user must still control SCF convergence and electronic-state continuity, especially during dissociation or radical encounters.

A user PES supplies a \code{PESCalculator} class with an energy function and either a force or a gradient function. An optional Hessian is used directly; otherwise central differences of gradients are available. This contract accommodates analytical surfaces, fitted polynomials, Fortran wrappers, and machine-learned models without requiring a particular fitting or training method. Calculators are reused to avoid repeated model loading. The repository includes several atom--diatom and atom--polyatom surfaces and a two-state Gaussian-process example for H$_2$O+Kr$^+$. Predictive variance is available for that particular model; uncertainty estimation is not a general requirement of the calculator interface.

Hessians for sampling and stationary-point work use the same backend layer. Analytical or backend-provided derivatives are used where supported; central finite differences provide a fallback. Stored Hessians carry geometry and electronic-model provenance, and mismatching files are not silently reused. A PES validation helper checks energy/force consistency against numerical differentiation. Such checks should precede dynamics on any new calculator, especially when wrapping a model with a different unit system.

Selected electronic backends can save orbital/wavefunction information in Molden format at output frames. This supplies data for external electronic-property analysis. It is distinct from the coordinate-based independent-atom scattering calculation described in Sec.~\ref{sec:scattering}, and no electronic wavefunction is available from a scalar fitted PES alone.

\subsection{Fewest-switches surface hopping}
\label{sec:fssh}

The optional nonadiabatic module propagates electronic amplitudes $c_i$ while the nuclei move on one active surface $a$, following the fewest-switches surface hopping (FSSH) framework \cite{Tully1990}. In an adiabatic representation,
\begin{equation}
 \dot c_i=-\frac{i}{\hbar}E_i c_i
          -\sum_j(\mathbf v\cdot\mathbf d_{ij})c_j,
 \label{eq:fssh}
\end{equation}
where $\mathbf d_{ij}$ denotes a derivative coupling. Electronic substeps use a unitary propagator for the generator held fixed during each nuclear step. This preserves electronic norm for that frozen generator; it does not remove the need to converge the nuclear timestep, electronic substeps, and hopping probabilities. Specifying the number of electronic substeps directly ensures that their durations span the nuclear step.

The present module approximates real coupling vectors from the adiabatic gap and its derivatives using a curvature construction associated with Baeck--An-type approaches and the Hessian-based treatment used in SchNarc \cite{Westermayr2020}. For a pair gap $\Delta E_{ij}=E_j-E_i$, the matrix used in the implemented construction reduces to
\begin{equation}
 \mathbf A_{ij}=\tfrac14\Delta E_{ij}\nabla\nabla\Delta E_{ij}.
\end{equation}
If its largest eigenvalue $\lambda_{\max}$ is positive, the estimated coupling is
\begin{equation}
 \mathbf d_{ij}\simeq\frac{\sqrt{\lambda_{\max}}}{|\Delta E_{ij}|}
                       \mathbf u_{\max}.
\end{equation}
Otherwise it is set to zero; a minimum-gap safeguard avoids singular division. The sign is chosen continuously relative to the previous step. Pairwise energy-gap cutoffs control where couplings are evaluated, and gradients and Hessians are skipped when no pair qualifies. Hops rescale momenta along the coupling direction subject to energy availability. An optional energy-based decoherence correction damps inactive-state amplitudes \cite{Granucci2007}, and state histories and electronic populations can be compared through ensemble consistency diagnostics.

The current state adapter calls PES calculators directly; general on-the-fly interfaces for nonadiabatic couplings or state overlaps from the electronic-structure backends are not implemented in this module. The curvature approximation is most naturally motivated near isolated two-state avoided crossings and requires independent validation for extended couplings, interacting multistate regions, or spin-orbit dynamics. Pairwise use for more than two surfaces does not establish its accuracy in a general multistate problem. The repository includes analytic Tully-model tests and comparison data to expose these limits. A minimum-energy crossing-point optimizer, described below, is a separate stationary-point tool and is not an intersystem-crossing dynamics algorithm.

\subsection{Photoionization preparation and ionic dynamics}
\label{sec:photoion}

The photoionization module maps a neutral phase-space ensemble onto one selected ionic PES at the same nuclear geometry, providing initial conditions for subsequent unimolecular ionic dynamics. It accepts either a supplied neutral reference geometry and Hessian, with the QCT prescriptions of Sec.~\ref{sec:qct}, or saved neutral MD coordinates and momenta. It does not generate the neutral MD or compute the source Hessian implicitly. Photon energy, electron kinetic energy, and ionic binding energy are external constraints; the latter two can be constants or tabulated distributions.

Let $V_n(\mathbf q)$ and $V_i(\mathbf q)$ be neutral and ionic potentials on compatible energy references, and let $\Delta V=V_i-V_n$ include any explicitly supplied offset. With photon energy $h\nu$, electron kinetic energy $\varepsilon$, and binding energy $E_b=h\nu-\varepsilon$, conservation gives
\begin{equation}
 K_i=K_n+h\nu-\varepsilon-\Delta V
    =K_n+E_b-\Delta V.
 \label{eq:photoenergy}
\end{equation}
Binding energy is therefore not an energy measured upward from the ionic minimum. Inconsistent or energetically inaccessible constraints are rejected. Neutral and ion surfaces with unrelated energy zeros require a physically justified offset.

Four preparation levels are implemented. Level~0 keeps the neutral momenta unchanged; its electron energy follows from the vertical gap. Level~1 scales all Cartesian momenta by $s=\sqrt{K_i/K_n}$ when this is feasible. This closes Eq.~\eqref{eq:photoenergy}, but also scales any pre-existing total linear and angular momenta. Level~2 instead separates mass-weighted momenta $\mathbf a=\mathbf M^{-1/2}\mathbf p$ into external and internal subspaces,
\begin{equation}
 \mathbf a=\mathbf a_{\mathrm{ext}}+\mathbf a_{\mathrm{int}},\qquad
 \mathbf a_i=\mathbf a_{\mathrm{ext}}+s\mathbf a_{\mathrm{int}}.
\end{equation}
Rescaling only the internal part preserves the specified total linear and angular momenta and assigns the required kinetic energy to internal motion. It cannot create a finite internal kinetic energy from an exactly motionless internal sample by scaling alone.

Level~3 additionally includes photon/photoelectron recoil,
\begin{equation}
 \Delta\mathbf p=\frac{h\nu}{c}\widehat{\mathbf k}
                 -\sqrt{2m_e\varepsilon}\,\widehat{\mathbf n}.
\end{equation}
The current electron-emission direction is isotropically sampled. Atomic weights $w_a$ distribute the recoil, $\Delta\mathbf p_a=w_a\Delta\mathbf p$, with $\sum_aw_a=1$. Center-of-mass weighting $w_a=m_a/M$ gives pure translational recoil; atom-localized or user-defined weights can add rotational and internal motion. These are explicit momentum-allocation models, not a calculation of ionization amplitudes or a unique reconstruction of a Dyson orbital.

Writing the internal recoil component as $\mathbf b$, Level~3 solves
\begin{equation}
 \tfrac12\|s\mathbf a_{\mathrm{int}}+\mathbf b\|^2=K_{i,\mathrm{int}}
\end{equation}
after the external recoil is accounted for. This retains recoil-induced changes in total momentum while satisfying the selected energy balance. The resulting ensembles describe conditional ionic launches; photoionization cross sections, vibronic transition intensities, and a detector response require additional physics and are not predicted by the launch module itself.

\subsection{Stationary-point, reaction-path, and thermochemical tools}
\label{sec:opt}

Minimum-energy optimization is available in Cartesian and redundant internal coordinates, with Hessian updates, step control, and convergence diagnostics. Transition-state searches use a partitioned rational-function optimization step that maximizes along a selected reaction mode while minimizing in the remaining subspace \cite{Baker1986}. Reaction-bond, atom-transfer, or internal-coordinate references can guide mode selection, and exact Hessian recalculation supports diagnosis near convergence. A transition structure must be characterized by its vibrational Hessian and connectivity, rather than inferred solely from optimizer termination.

RDA-based quasi-transition-state guesses are constructed from reactant and product endpoint geometries. The module also provides optional chain refinement and path profiles. These guesses can be passed to a dedicated TS search; a high point on an approximate path is not itself proof of a first-order saddle point. IRC following traces both directions from an imaginary mode, with mass-weighted stepping, correction, and optional endpoint minimization. One- and two-dimensional relaxed scans, normal-mode scans, and evaluation of XYZ geometry sequences complement the optimization tools.

The spin-crossing optimizer combines energies and gradients from two selected spin surfaces to approach a minimum-energy crossing point. Convergence requires both a small energy gap and the appropriate projected-gradient conditions. This identifies a crossing geometry and energy; a spin-transition probability additionally requires coupling and dynamical information.

Normal-mode analysis provides frequencies, displacement animations, and thermochemistry. The latter includes rigid-rotor/harmonic-oscillator contributions and an optional Grimme-type quasi-RRHO entropy treatment for low-frequency modes \cite{Grimme2012}. Temperature, pressure, rotational symmetry number, and chirality factor are explicit inputs. The quasi-RRHO entropy correction does not replace the physical frequencies in the QCT sampler and is not a molecular dynamics thermostat.

\subsection{Parallel execution, restart records, and graphical input}
\label{sec:scratch}

Independent trajectories are scheduled with separate output and scratch directories and a controllable number of electronic-structure threads per job. Seed derivation from trajectory indices makes sampling reproducible with respect to scheduling. Per-trajectory status records provide progress and failure information without requiring all workers to finish successfully. Because electronic-structure parallelism and ensemble parallelism coexist, their product should be matched to the available resources.

Checkpoints include coordinates and momenta, with sidecar state for random generators, thermostats, integrator history where applicable, constraint metadata, and reaction-channel persistence. Electronic-state restoration also requires amplitudes, active-state identity, and consistent surface definitions. Exact continuation is meaningful only with compatible model parameters and complete state; a geometry alone is not a reproducible restart. The Qt-based graphical input builder generates editable scripts for common unimolecular, collision, and optimization tasks. The Python API remains the complete interface for advanced preparation and analysis.

\section{Analysis of trajectory ensembles}
\label{sec:analysis}

\subsection{Product identity, energy partitioning, and state assignment}
\label{sec:final}

The post-collision routines identify connected product fragments using geometric connectivity and evaluate their momenta and energies in their respective center-of-mass frames. The graph thresholds and named channel definitions should be chosen consistently. When exactly two detached products are present, relative translation, orbital angular momentum, and the associated scattering geometry are defined. For a different number of products, the routines retain per-fragment and conservation information rather than misassigning two-body observables.

For product $A$ with internal Cartesian momenta $\widetilde{\mathbf p}_a$, the internal kinetic energy and angular momentum are
\begin{equation}
 K_{A,\mathrm{int}}=\sum_{a\in A}\frac{|\widetilde{\mathbf p}_a|^2}{2m_a},
 \qquad\mathbf J_A=\sum_{a\in A}\mathbf r_a\times\widetilde{\mathbf p}_a.
\end{equation}
A rotational contribution is assigned as
\begin{equation}
 E_{A,\mathrm{rot}}=\tfrac12\mathbf J_A^T\mathbf I_A^+\mathbf J_A.
\end{equation}
The inertia can be evaluated at the instantaneous geometry or at a supplied equilibrium structure aligned by a mass-weighted best fit. If both instantaneous and reference isolated-product potentials are available,
\begin{equation}
 E_{A,\mathrm{vib}}=K_{A,\mathrm{int}}-E_{A,\mathrm{rot}}
                      +V_A(\mathbf q_A)-V_A(\mathbf q_{A,\mathrm{eq}}).
 \label{eq:productvib}
\end{equation}
The code does not equate residual kinetic energy with full vibrational energy when those potential references are unavailable. Electronic charge and spin assignments for products must also be specified consistently. At finite product separation, interfragment interaction energy can prevent an asymptotic energy partition, so separation convergence and conservation residuals are useful diagnostics.

State-resolved interpretation requires a further mapping from these classical data to semiclassical actions \cite{Hase1998final,Peslherbe1999}. For a nearly harmonic diatom, $n^{\mathrm{eff}}=E_v/(\hbar\omega)-1/2$ and $J^{\mathrm{eff}}=[\sqrt{1+4|\mathbf J|^2/\hbar^2}-1]/2$ are possible continuous labels; anharmonic products require a corresponding action model. Polyatomic mode projection additionally requires an atom mapping, equilibrium Hessian, and consistent rotational alignment. Histogram or Gaussian binning can then be applied in external ensemble analysis \cite{BonnetRayez1997,BonnetRayez2004}. The present implementation provides the phase points and energy diagnostics, but a general automated product-action assignment and Gaussian-binning pipeline is not part of the analysis interface. The results below therefore use a generic channel indicator or weight that the ensemble analysis must define explicitly.

\subsection{Cross sections and thermal rate estimators}
\label{sec:rates}

Let $h_{ir}$ be one when trajectory $i$ reaches channel $r$ and zero otherwise. More generally it can represent a specified final-state weight. For a sampling density $g(b)$ normalized on $[0,\bmax]$, the cross-section estimator at fixed collision energy and specified reactant preparation is
\begin{equation}
 \widehat\sigma_r=\frac{1}{N}\sum_{i=1}^N
                 \frac{2\pi b_i}{g(b_i)}h_{ir}.
 \label{eq:sigmageneral}
\end{equation}
For uniform-area sampling, $g(b)=2b/\bmax^2$, this becomes
\begin{equation}
 \widehat\sigma_r=\pi\bmax^2\frac{N_r}{N},
 \label{eq:sigma}
\end{equation}
whereas uniform-$b$ sampling gives
\begin{equation}
 \widehat\sigma_r=\frac{2\pi\bmax}{N}\sum_i b_i h_{ir}.
\end{equation}
The impact-parameter range must cover the reactive region. The equations assume a common $\bmax$; energy-dependent or channel-adaptive sampling requires retaining the actual sampling density for each draw.

Thermal averaging can be written
\begin{equation}
 k_r(T)=\overline v(T)\int_0^\infty p_{\mathrm{flux}}(E)\,
                        \sigma_r(E)\,dE,
 \quad \overline v=\sqrt{\frac{8\kb T}{\pi\mu}}.
\end{equation}
Consequently, with collision energies drawn from Eq.~\eqref{eq:ecolltherm},
\begin{equation}
 \widehat k_r=\frac{\overline v}{N}\sum_i
                     \frac{2\pi b_i}{g(b_i)}h_{ir}.
 \label{eq:kT}
\end{equation}
If the internal states are also sampled thermally at $T$, this estimates the corresponding thermal rate within the preparation model. With selected internal states it estimates a translationally averaged state-selected rate. Cross sections at fixed energy and opacity functions at fixed $b$ instead require ensembles prepared at those conditions. Statistical electronic weights, when required, must be applied for the explicitly represented electronic surfaces.

For independent draws, define $X_i=2\pi b_i h_{ir}/g(b_i)$. An estimated standard error is
\begin{equation}
 \mathrm{SE}(\widehat\sigma_r)=
 \left[\frac{\sum_i(X_i-\overline X)^2}{N(N-1)}\right]^{1/2},
 \label{eq:weightederror}
\end{equation}
and is multiplied by $\overline v$ for Eq.~\eqref{eq:kT}. A zero observed count does not imply a vanishing uncertainty or an exactly zero rate; binomial confidence bounds are appropriate in that limit. Correlated source MD frames require an effective-sample-size or block analysis. Weighted channel branching ratios use ratios of the corresponding weighted channel sums. These equations specify post-processing of the recorded ensemble; the current code does not automatically accumulate every cross section, rate, opacity, or state-to-state distribution named here.

\subsection{Energy transfer and stereodynamical correlations}
\label{sec:et}
\label{sec:stereo}

For a nonreactive encounter, the initial and final internal energies determine $\Delta E_A=E_{A,\mathrm{int}}'-E_{A,\mathrm{int}}$. Weighted histograms and moments can then be constructed for comparison with energy-transfer models \cite{Lendvay2019,OrefTardy1990,Barker2001,Troe1977}. For example, a cross-section moment is
\begin{equation}
 \widehat\Sigma_m=\frac1N\sum_i
     \frac{2\pi b_i}{g(b_i)}h_{i,\mathrm{nr}}(\Delta E_{A,i})^m,
\end{equation}
and the conditional mean over the selected nonreactive collision class is $\widehat\Sigma_1/\widehat\Sigma_0$. Defining that class is important: adding arbitrarily distant, negligibly perturbing encounters changes a per-collision mean. Conversion to a master-equation transfer kernel additionally requires the energy-grain definition, collision normalization, and ensemble convention. Those ensemble constructions are external post-processing of the available product and reactant energy records.

The built-in vector analysis records initial and final relative velocities, orbital angular momenta, fragment rotational angular momenta, scattering angles, and intervector correlations. These quantities support stereodynamical distributions and differential cross sections after ensemble weighting. Isotropic rotation is the default preparation; specially aligned ``helicopter,'' ``propeller,'' or ``cartwheel'' ensembles require explicit custom phase points and are not advertised as dedicated high-level sampling switches. Likewise, a reaction termination time is not automatically a Smith collision-delay observable \cite{Smith1960}: extracting a residence time or delay requires a specified interaction region or asymptotic flight-time reference.

\subsection{Velocity spectra and fragment-resolved time--frequency analysis}
\label{sec:spectra}

The basic vibrational analysis computes a velocity power spectrum,
\begin{equation}
 S_v(\omega)\propto\sum_{a\alpha}
      \left|\sum_n v_{a\alpha}(t_n)e^{-i\omega t_n}\right|^2.
 \label{eq:velspec}
\end{equation}
Its frequency content is related to a velocity-autocorrelation spectrum under the corresponding Fourier conventions \cite{Thomas2013}. This signal identifies nuclear motions but is not, by itself, an infrared or Raman intensity: those require dipole or polarizability observables and their appropriate correlations.

The short-time analysis applies a moving window,
\begin{equation}
 S_v(t,\omega)\propto\sum_{a\alpha}
 \left|\sum_n w(t_n-t)v_{a\alpha}^{\mathrm{int}}(t_n)
                        e^{-i\omega t_n}\right|^2,
 \label{eq:stft}
\end{equation}
with options to project out translation and rotation and to use mass-weighted internal velocities. Window duration trades temporal localization against frequency resolution. Dynamic fragment tracking uses connectivity histories to associate signals with emerging fragments, enabling channel- and fragment-resolved inspection of an adduct, rearrangement, or dissociation. Connectivity thresholds and persistence settings are part of that analysis definition. The time interval in Eqs.~\eqref{eq:velspec}--\eqref{eq:stft} is the actual stored-frame spacing, which need not equal the integration timestep.

\subsection{Structural form-factors from X-ray and electron scattering}
\label{sec:scattering}

The scattering module evaluates rotationally averaged independent-atom-model (IAM) signals directly from trajectory coordinates. For momentum-transfer magnitude $Q$ and interatomic separation $r_{ab}(t)$, its elastic component has the Debye form
\begin{equation}
 I_{\mathrm{el}}(Q,t)=\sum_{a,b}f_a(Q)f_b(Q)
              \frac{\sin[Qr_{ab}(t)]}{Qr_{ab}(t)},
 \label{eq:iam}
\end{equation}
where the self-term limit is one. The atomic incoherent functions supply $I_{\mathrm{inel}}(Q)=\sum_a S_a(Q)$ and $I_{\mathrm{tot}}=I_{\mathrm{el}}+I_{\mathrm{inel}}$ within this model \cite{Hubbell1975}. The code uses tabulated elastic coefficients and incoherent functions distributed with the program, with the conversion between $Q$ and the tabulation variable $s=Q/(4\pi)$ made explicitly. Coverage follows the available atomic tables rather than a restriction to H, C, N, and O.

Temporal Fourier transforms resolve the vibrational frequencies modulating a difference-scattering signal; short-time transforms localize those changes along the trajectory. A sine transform of a suitably normalized momentum-transfer difference gives a pair-distribution-type real-space signal, whose resolution and oscillatory artifacts depend on the finite $Q$ range and damping. For fixed atomic composition, the IAM incoherent sum has no structural time dependence, so it does not acquire molecular vibrations merely by following the trajectory.

The IAM calculation does not use the saved Molden wavefunctions. It therefore cannot describe changes in bonding electron density, coherent electronic-state superpositions, or general electronic inelastic transitions beyond its atomic approximation. Wavefunctions saved by the electronic backends can support a separate, more detailed scattering calculation, including applications to ultrafast x-ray or electron diffraction \cite{Minitti2015,Yang2018}, but this is an extension through external analysis rather than an already implemented wavefunction-based electron-scattering module.

\section{Validation and test systems}
\label{sec:benchmarks}

\subsection{Sampling and propagation validation}

Initial-condition validation should first compare the sampled distributions with their analytical targets. Fixed-energy harmonic draws must reproduce each requested mode energy before rotational adjustment. Thermal harmonic populations should reproduce the geometric distribution and Eq.~\eqref{eq:thermalmean}; Wigner draws should reproduce their coordinate and momentum variances and distributed harmonic energies. Thermal diatomic $J$ histograms should be compared with Eq.~\eqref{eq:diatomrot}, while general-top rotation should reproduce $\langle J_\alpha^2\rangle=I_\alpha\kb T$ and the appropriate equipartition energy. Linear rotors provide a necessary test of the excluded axial rotation. Rotating-Morse ensembles should be checked against the bounded joint distribution in Eq.~\eqref{eq:morsecoupled}.

The next level tests assembled Cartesian phase points: zero center-of-mass momentum, realized total angular momentum, rotational invariance, rigid constraints where selected, and the change between harmonic preparation energy and the full-PES initial energy. Production tests should report timestep convergence, energy and momentum conservation, restart continuity, and sensitivity to reaction-channel thresholds. These checks are represented in the repository's automated test suite, but passing software tests is distinct from validation of a potential model or of a chemical prediction.

\subsection{Reaction, preparation, and observable comparisons}

Analytical-PES calculations for atom--diatom and atom--polyatom reactions provide inexpensive tests of excitation functions, impact-parameter convergence, and vibrational-state effects. H+O$_2$ on the available electronic-state surfaces is a natural connection to established trajectory literature \cite{Xu2005,Li2010,SzaboLendvay2015a,SzaboLendvay2015b}, and the same preparation and analysis apply with on-the-fly forces to direct-dynamics radical--O$_2$ capture \cite{Szabo2025capture}. Any such calculation is specified by the source commit, PES or electronic-structure version, reactant preparation, number of trajectories, stopping rules, and weighted statistical uncertainties.

Separate module demonstrations can show constrained versus flexible collision partners, neutral-to-ion energy and momentum conservation for all photoionization levels, and the response of velocity and scattering spectra to an excited normal mode. For the nonadiabatic module, the analytic Tully models \cite{Tully1990} included with the code allow exact and approximate couplings to be compared, and electronic populations and trajectory branching to be followed while the nuclear and electronic timesteps are converged; such comparisons distinguish coupling-model errors from propagation and hopping errors.

\section{Conclusions}
\label{sec:conclusions}

\Smite\ combines mode-resolved reactant preparation, classical propagation, and trajectory analysis with stationary-point and reaction-path tools that share the same energy and derivative interfaces. Its principal methodological flexibility is the ability to choose vibrational, rotational, and relative-motion ensembles independently while retaining explicit control over their physical meaning. Harmonic fixed-action and Wigner sampling, coupled rotating-Morse preparation, saved MD phase points, and photoionization launches cover distinct experimental and theoretical initial-condition problems.

The modular implementation connects these preparations to on-the-fly electronic structure or analytical and machine-learned PESs, with optional rigid constraints, thermalization, and PES-based surface hopping. Product energy partitioning, vector correlations, short-time vibrational analysis, and coordinate-based x-ray scattering provide complementary views of the resulting dynamics. Quantitative use requires convergence and validation of each preparation, potential, propagator, and observable; the explicit formulas and module boundaries described here make those choices traceable and provide a basis for further development.

\section*{Code availability}

The source code, examples, and tests are available at
\url{https://github.com/peter88szabo/Smite} under the repository's GNU General Public License.

\begin{acknowledgments}
This research was supported by the Belgian Federal Science Policy Office under grant Prf-2019-052\_REVOCS of the FED-tWIN programme.
\end{acknowledgments}


\end{document}